\documentclass[smallextended]{svjour3}       
\usepackage{rotating}
\usepackage{multirow} 
\usepackage{amstext}
\usepackage{amssymb}
\usepackage{graphicx}
\usepackage{amsmath}
\usepackage{multicol} 
\usepackage{listings}
\usepackage{pdfpages}
\usepackage{subcaption}
\usepackage{array}

\smartqed  
\usepackage{graphicx}
\begin{document}

\title{ Characterizing Communicability of Networks formed on Mobile Nodes }

\author{Karim Keramat Jahromi}

\institute{{Karim Keramat Jahromi \at
		Computer Science Department, University of Milan}, Italy\\
	\email{karim.keramat@unimi.it} 
	}

\maketitle

\begin{abstract}
	
Smartphones have become extremely popular by launching wide ubiquitous networks. Nowadays studying of DTN Delay Tolerant Networks (DTN) and Opportunistic Networks where formed over these mobile nodes, is one of the interesting topics in the research community. In this paper, we measure communicability capacity of DTN Network formed over the mobile nodes at a university campus and also an area in Montreal city through exploiting static and temporal graphs. We observed a significant difference between communicability measures in static and temporal cases, especially for short snapshot windows. It implies that analyzing dynamic networks by considering a static model for them may lead to an unrealistic and even mislead results.


\end{abstract}

\section{Introduction}\label{sec_intro}

 Complex network is an area of science that analyzes the behavior of any network through graph properties and tries to relate them to the real aspects of these networks. The identified properties are not trivially observed and they are usually the result of recurrent behaviors in the network. By this manner, the studying of networks through the complex networks’ perspective provide insightful information to a deeper understanding of their characteristics. Nowadays large-scale complex networks arise in a range of natural and technological fields, from biological to the telecommunication system, and they impose many challenges and computational mathematicians.
  DTN and opportunistic networks (networks that maintain communication capabilities even within regions with no previously deployed infrastructure communication in a multi-hop fashion between mutually reachable devices) \cite{Huang2008} behave distinctly from other networks and a useful way to study and determine their intrinsic attributes, is from the perspective of complex networks. When we observe the  opportunistic networks as complex networks, we can identify non-trivial fundamental aspects by the analysis of the attributes of the formed complex networks. Therefore, we are able to understand clearer how these networks behave and propose more suitable solutions.
  Social network analysis is one of the fields that so many key ideas in complex network area arise where million or more of mobile nodes interact with each other.
  
  In this paper, we provide an extensive analysis of the structural properties of contacts among mobile nodes which form the DTN and opportunistic networks. To do so, we represent contacts and interaction among mobile nodes in a compact and tractable way in two cases as a weighted contact graph in static (aggregated) graph and  also an unweighted contact graph in Temporal network. 
  In a static weighted contact graph, the weights (i.e., tie strengths) express how frequently and how long pairs of mobile nodes are in contact.
 Given such a contact graph, we can use tools and metrics from social network analysis and graph theory (e.g., Centrality metrics, Communicability, etc.) to quantify the amount of structure in the underlying mobility datasets. 
 Furthermore, it is important to achieve an estimation of communication capacity in DTN and opportunistic networks to get an approximation of speed and delay in diffusion of data over these kinds of networks.
 Our main findings and contributions can be summarized as follows: 
 first of all, we proposed a new measure for strength estimation  of social tie among mobile nodes by considering Inter Contact Time (ICT) alongside the Contact Duration Time (CT) and also frequency of Contact which is more realistic compared to other previously proposed definitions. 
 Also, we characterize and compare the empirical distributions and average of total communicability per edges/nodes and centrality for weighted and unweighted graph networks with different sizes (number of nodes) for real-world datasets (university campus and Montreal city WiFi datasets and also in larger scale the smartphones dataset ) and also the synthetic networks. We also calculate and compare the temporal communicability capacity through exploiting different time windows (snapshots).

\section{Related Work}\label{sec_RelatedWork}

Nowadays smartphones have become more and more ubiquitous and popular, and analyzing wireless networks formed over these devices is becoming an important research field. Since encounter events among smartphones (mobile nodes) provide the communication opportunities in DTN and opportunistic networks, knowledge about the encounter patterns, their characteristics and capacity of message communications that affect the speed and delay in network, is important in opportunistic Networks. On the other hand studying of networks formed over encountered nodes, through the complex networks’ perspective provides the insightful information to a deeper understanding of their characteristics. Hossmann et al. \cite{Hossmann2011} used complex network for analyzing contact characteristic such as community detection, node degree among mobile nodes by defining social weighted graph. Authors defined the tie-strength of weighted graph according to the contact duration and frequency of contacts occur among encountered nodes. In this work for definition of social weight addition to the contact duration and frequency of contact, also the Inter Contact time has been considered that make more realistic sense of social weight among encounters. Benzi et al. \cite{Benzi2013} proposed several metrics for measuring the importance of node in unweighted static networks which is called node centrality. Crofts et al. \cite{Crofts2009} used a weighted communicability measure applied to complex brain networks to distinguishing local and global differences between disease patient and control. Tang et al. \cite{Nicosia2013} proposed new temporal metrics to quantify and compare the speed of information diffusion by considering evolution of network over time. The proposed temporal metrics capture the characteristics of time-varying graphs such as delay, duration and time order of contacts. Authors in \cite{Estrada2012,Santoro2011} proposed different local and global temporal centrality metrics such as Closeness, Betweenness, Eigenvalues and Katz for analyzing the structure of social networks. Here we propose a new metrics for measuring social strength tie among encounters in static network. We evaluated and compared average of centrality and communicability per nodes/edges among different size of synthetic and realistic datasets based networks by modeling these network with static graph. Also we calculated total communicability for temporal network over different snapshot windows, so high dependency on size of snapshot window was observed.  

\section{Datasets}\label{sec_Dataset}

In our experiment, we used two WiFi datasets of smartphones each covering a different mobility scenario. WiFi datasets are extracted from logs of the RADIUS service, from a university campus in Europe and another one from several WiFi hotspots in Montreal city in Canada \cite{MontralDataset}. Whenever a station (smartphone, tablet or laptop computer) associates or disassociates with an AP (Access Points), a Syslog message is recorded. Each record contains a time stamp in seconds, the MAC addresses of the AP and the mobile node, the Access Session Time in seconds, and the Access Session Status (Start - association), or (Stop - disassociation). The analyzed university campus dataset is 5 months long from 1 Jan 2011 to 31 May 2011. This university campus dataset contains references to 1381 APs and 18137 mobile nodes. The Montreal dataset is 3 years long from 28 August 2004 to 28 August 2007 and contains 206 APs and 69689 mobile nodes. Both WiFi datasets do not include any information about the geographical coordinates of the APs and their spatial distributions. We try to extract encounter events among mobile nodes by performing some preprocessing to remove ping-pong events and extracting pair encounters nodes \cite{Keramat2014}.


\section{Encounter Events}\label{sec_EncounterEvent}

An encounter event means meeting face to face, which implies physical proximity among people. The extent of this physical proximity is not always exactly clear and may be vary on different scenarios, applications, and domains. For instance, in the biological field and in disease spreading, physical proximity is short, while in wireless networks it depends on the coverage areas of mobile devices or wireless network infrastructures. Nowadays smartphones are so widely carried by humans that can be used to observe mobility and extract physical proximity information. In the communication network literature, an encounter among mobile devices occurs when they are in the communication range or when they are within the same coverage area of the communication network infrastructure, the latter also called indirect encounter \cite{Legendre2011}. Although this definition may not always reflect proper and exact realistic physical encounters among mobile nodes due to some challenges \cite{Jahromi2014,Jahromi2015}, 
most researchers define an encounter event occurrence in a WLAN when two or more mobile nodes are associated to the same AP during an overlapped time interval. 
Despite some challenges and limitations, if collected WiFi datasets are used carefully (i.e. accounting for the effects of ping-pong events, overlap in coverage areas and missed encounters) it would appear to be a rich source of empirically-derived data on human encounters since large amount of data can be gathered easily at low cost, allowing even large-scale analysis of encounter patterns.
Here for WiFi dataset, we use smoothing the ping-pong events according to \cite{Jahromi2014}, for extracting encounter events.

The resulting record for an encounter event is:
\begin{verbatim}
UserA,UserB,PoI Id,Encounter Start Time,Encounter End Time.
\end{verbatim}

\section{Static Network}\label{sec_Static Network}

Real-world networks are usually modeled by means of graphs. In this section, the characteristic of Static (aggregated) social weighted Graph and some metrics for analyzing node centrality and communicability in static graph network will be discussed.

\subsection{Characterizing Static Social Network Graph}\label{sec_Static Graph}

A graph $G=\left (V,\right. E) $ is a set of nodes $V$ as vertices with cardinality $n$, and edges  $E=\left \{ (i,j)\mid i,j\epsilon V \right. \} $.

A graph is undirected if the edges are unordered among pairs of vertices. Each graph could be represented as a matrix which called adjacency matrix. The Adjacency Matrix of a network with graph $G$ is given by

\begin{equation} \label{eq.1}
	\begin{array}{l}
a_{ij}=\left\{\begin{matrix}
	1& if   \left ( i \right. j)  \varepsilon  G \\ 
	0& else
 \end{matrix}\right.
	\end{array}
\end{equation}

 Self-link (loop) is not allowed; $a_{ii}=0$ and for undirected graph $ a_{ij}=a_{ji} $ is satisfied. 
 When the network is undirected, the adjacency matrix will be symmetric and eigenvalues of $A$ will be real. A graph $G$ is weighted if the numerical values are associated with its edges. If all the edges are given the same value 1, we say that the graph is Unweighted. 

\subsection{Social Weight Graph}

One of the classic ways of representing social networks is by weighted graphs. In this type of graph, each vertex represents one mobile node, and an edge exists between two nodes if at least one encounter event (for WiFi dataset) or colocation event ( for CDR dataset ) has been detected between them. The weight associated with each edge of the graph is used to model the strength of the interactions between pairs of nodes. Different metrics have been proposed for measuring social ties, such as Aggregate contact duration, the frequency of contact \cite{Hui2011} or even combination of these two metrics \cite{Hossmann2011}. 

In the proposed approach the social strength between encountered/colocated mobile nodes can be estimated based on the average Contact Time Duration and Inter Contact Time  \cite{Chaintreau2007}  and also the frequency of encounter events. Encounter/Colocation Duration Time (CT) specifies for how long two nodes are in direct contact with each other (or how long their encounter/colocation event takes) and Inter Contact Time (ICT) is defined by the time interval elapsed between two consecutive contacts ( encounter or colocation event) of the same pair of nodes, it determines how often an encounter or colocation event among nodes is possible. A shorter average ICT means that two people meet each other quite often. If the ICT of two nodes is long, that means they have to wait a long time for the next encounter/colocation event. It can imply that the CT and ICT capture the closeness between encountered and colocated nodes. Although such closeness may or may not reflect actual friendship in a social context, it indicates the relationship between wireless devices as revealed in their association patterns. Then social weights between encountered/colocated nodes can be described by the statistics of CT and ICT. It means that higher CT and shorter ICT implied a simply stronger relationship between mobile nodes and vice versa. Another factor that should be considered in calculating social weights is the frequency of encounter/colocated events (number of times that encounter/colocation event occurs between a specific pair of nodes within a certain observation period). It is logical that friends meet each other more frequently than acquaintance and strangers. For instance, one node may have long CT duration with another one, followed by a long ICT interval. In another case, the node may have a short CT followed by a short ICT. So it is necessary that the frequency of encounters/colocation is considered in calculating the social weight otherwise, there is no difference between the calculated social weights in the two mentioned cases in above.

According to the above discussion, by considering $N_{i}$ and $N_{j}$ as total number of encounter/colocation events occurrence for node $i$ and node $j$, respectively, $n_{ij}$ as the number of times that encounter/colocation events occurs between node $i$ and node $j$, $t_{CT}^{k}(i,j)$ as encounter/colocation duration between node $i$ and $j$ in $k^{th}$ occurrences and  $t_{ICT}^{p}(i,j)$ as the ICT between node $i$ and $j$ in its $p^{th}$ occurrences, then the social weight between node $i$ and $j$ can be defined as:

\begin{equation} \label{eq.2}
		\begin{array}{l}
	W_{ij}=\frac{\overline{T_{CT}(i,j)}}{\overline{T_{ICT}(i,j)}} \left [ \frac{2n_{ij}}{N_{i}+N_{j}} \right ]
       \end{array}
\end{equation}

Where $\overline{T_{CT}(i,j)}$ and $\overline{T_{ICT}(i,j)}$ are respectively the average encounter/colocation and average ICT between node $i$ and $j$ over the entire observation period of dataset, calculated according to the below equations:

\begin{equation} \label{eq.3}
	\begin{array}{l}
	
	\overline{T_{CT}(i,j)}=\frac{1}{n_{ij}} \sum_{k=1}^{n_{ij}}t_{CT}^{k}(i,j)
	
	\end{array}
\end{equation}

\begin{equation} \label{eq.4}
	\begin{array}{l}
		
			\overline{T_{ICT}(i,j)}=\frac{1}{n_{ij}-1} \sum_{p=1}^{n_{ij}-1}t_{ICT}^{p}(i,j)
		
	\end{array}
\end{equation}

Here we use the same value to represent the interaction between two encontered/colocated nodes, i.e., it is assumed that the social strength between node $i$ and $j$ is equal to the social strength between node $j$ and $i$. It means that social graph is undirected. 

\begin{figure}
	\centering
	\includegraphics[width=1\textwidth]{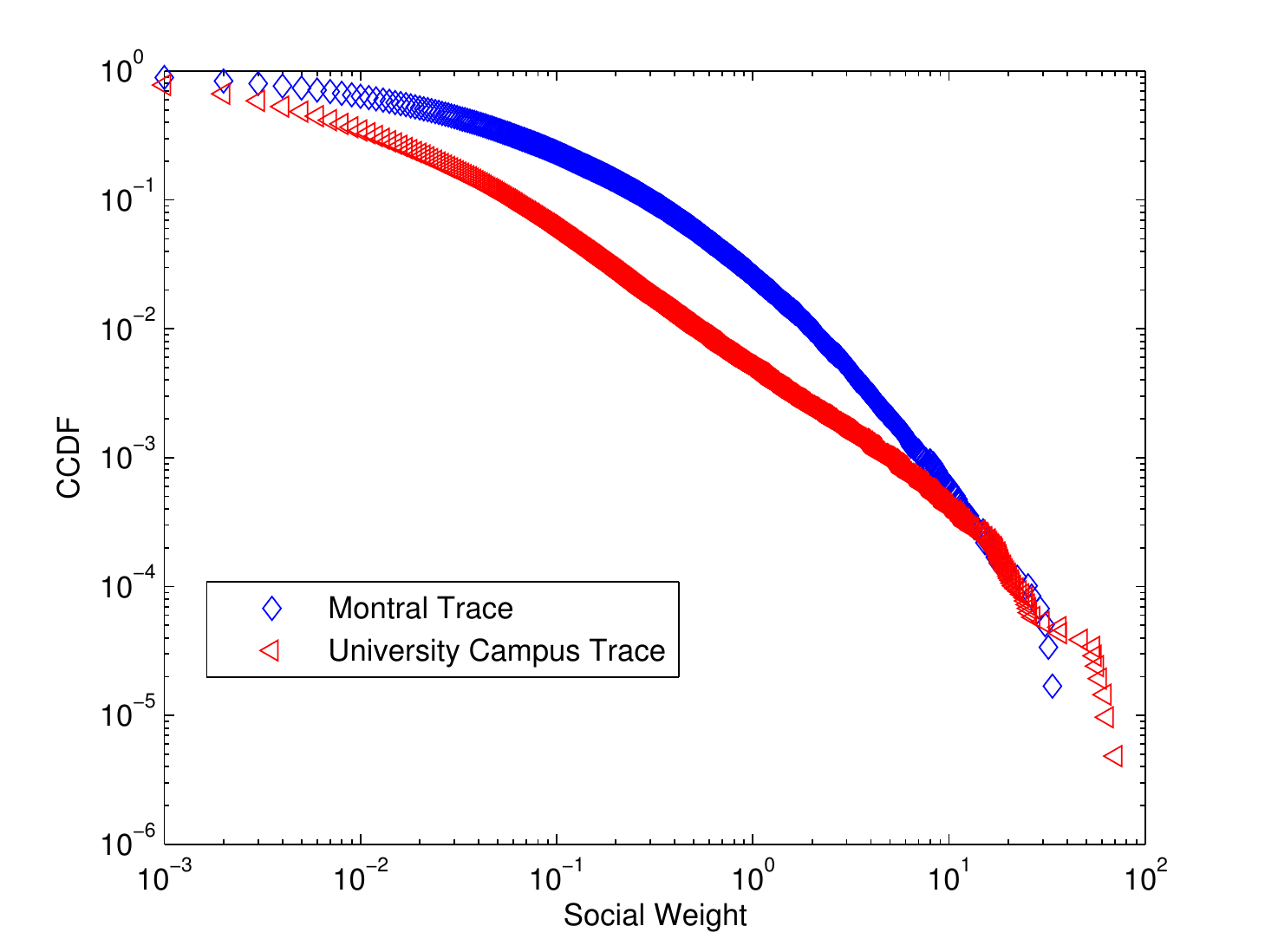}
	\caption{The empirical distributions of social weight follow Heavy-Tail trend.}
	\label{fig:socialweight}
\end{figure}

The empirical distributions of the social weight values of encounter events on the university campus and Montreal city datasets follow Heavy tail trends as shown in Figure \ref{fig:socialweight}. Especially for a university campus, the distribution depicts almost Power-Law trend. This implies that the majority of pair nodes do not have tight relationships together, although they do encounter. We can conclude that social relationships between mobile nodes are sparse.

Although authors in \cite{Zhang2015}, by analyzing MIT Reality Mining CDR dataset, have observed a strong correlation between calling pattern and colocation patterns of mobile users, nowadays the majority of contact actives among people and friends have been oriented towards the wide variety of Internet-based applications. However, such Internet-based contact activities somehow will be hidden from CDR datasets, since when a user accesses to the Internet just the Internet traffic data will be recorded in CDR. Therefore nowadays we should be  conservative about retrieving the social interplay among mobile users just relying on the CDR datasets.

\begin{figure}
	\centering
	\includegraphics[width=1\textwidth]{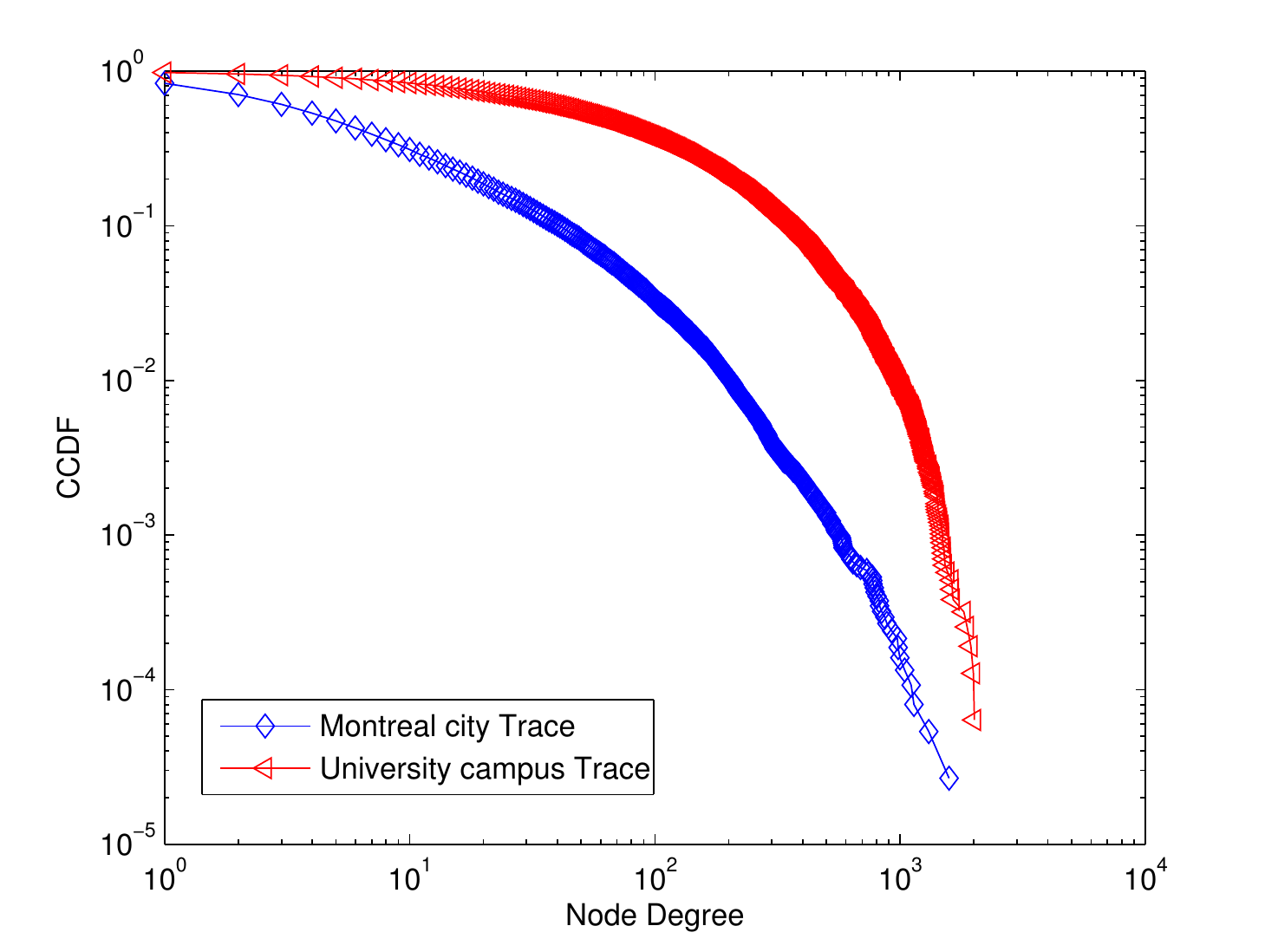}
	\caption{The empirical Node Degree distribution follow Heavy-Tail trend.}
	\label{fig:NodeDegree}
\end{figure}

\subsection{Analyzing Centrality}

Measures of node importance are usually referred to node centrality, means understanding which one is the most central node that could help disseminating the message in the network in a faster way. 
Many centrality measures have been defined \cite{Benzi2013} starting with simplest such as node degree. But node degree which is the number of distinct nodes directly connected (encountered/colocated) with a specific node, just retrieves local information not connectivity of the immediate neighbors of the node under study, so the importance of an adjacent node is not considered. Distributions of node degree for both datasets have been shown in Figure \ref{fig:NodeDegree}. Distributions follow Heavy-Tail trend which implies the majority of nodes have small node degree and a minority of nodes have very high node degree. 

Other popular notations of centrality are betweenness and closeness centrality \cite{Benzi2010} which is defined according to the shortest path which again just retrieve local information. These two centrality metrics assume that all communication in the network takes place via shortest path but this is often not the case.
In order to take into account the global structure of the network and the fact that all walks between all the pairs of nodes should be considered not just shortest one, \textbf{sub-graph centrality} have been proposed \cite{Benzi2010}. Sub-graph centrality measures the centrality of a node by taking into account the number of sub-graphs that the node participates in. The smaller sub-graphs are given more weight than larger ones. 

The sub-graph centrality  of node $i$ is given by $\left [e^{A} \right ]_{ii}$ (entity in $i^{th}$ row and $i^{th}$ column of matrix $e^{A}$), where $A$ is the unweighted adjacency matrix of the network graph \cite{Chung2006}. The node has large sub-graph centrality is considered to be more important in the network and is given a higher ranking than nodes with lower sub-graph centrality \cite{Benzi2014}.

A walk of length $k$ on a graph $G$ is a sequence of vertices $v_{1},v_{2},\cdots, v_{k+1}$ such that $(v_{i},v_{i+1})\in E$ for all $1\leq i \leq k$. A path is a walk with no repeated vertices. A closed walk is a walk that starts and ends at the same vertice. In graph theory if $A$ is the adjacency matrix of a network with unweighted edge then $\left [A^{k} \right ]_{ij}$ counts the number of walks of length $k$ between nodes $i$ and $j$ \cite{Chung2006}.
So $\left [ e^{A} \right ]_{ii}$ that is the entity of $i$ th row and $i$ th column of Matrix  $e^{A}$, and is called  sub-graph centrality of node, is equal to the counts the number of closed walks centered at node $i$ weighting a walk of length $k$ by a penalty factor of $\frac{1}{k!}$ since short walks are more important than long ones. 

In message passing scenario, shorter walks are faster and cheaper as result walks by length k are penalized by weight $\frac{1}{k!}$.
Note that according to the linear algebra

\begin{equation} \label{eq.5}
\begin{array}{l}

e^{A}\cong I+A+\frac{A^{2}}{2!}+\frac{A^{3}}{3!}+\cdots +\frac{A^{k}}{k!}+\cdots =\sum_{k=0}^{\infty} \frac{A^{k}}{k!}

\end{array}
\end{equation}

And $\sum_{j=1}^{n}\left [e^{A} \right]_{ij}$
that is the row sum of $e^{A}$ for node $i$, counts all walks between node $i$ and all the nodes in the network, weighting walks of length $k$ by the penalty factor of $\frac{A^{k}}{k!}$.
   
For graph network with weighted edges, such as social weight graph, the number of walks is weighted by product of weight of each edge. For the weighted adjacency Matrix $A$ and also Diagonal matrix $D$ where \\ 
$D=diag(d_{1},d_{2},\cdots,d_{n})$  is the degree matrix and $d_{i}=\sum_{k=1}^{N} a_{ik}$; where $a_{ik}$ is entity of $i$ th row and $k$ th column of adjacency matrix $A$ and $N$ is number of nodes in the weighted graph.
The sub-graph centrality of node $i$ is calculated as

\begin{equation} \label{eq.6}
\begin{array}{l}

S_{i}=\left [ exp(D^{\frac{-1}{2}}AD^{\frac{-1}{2}}) \right ]_{ii}

\end{array}
\end{equation}

And the Normalized Total sub-graph centrality is defined as $\frac{1}{N}\sum_{i=1}^{N} S_{i}$. 

\subsection{Analyzing Communicability}

Communicability measures how is easy to send a message from node $i$ to node $j$ in a graph \cite{Estrada2007,Estrada2012}.
For the case that adjacency matrix $A$ is weighted, the communicability between distinct node $i$ and $j$ in a weighted graph is calculated by this formula:

\begin{equation} \label{eq.7}
\begin{array}{l}

C(i,j)=\left [ exp(D^{\frac{-1}{2}}AD^{\frac{-1}{2}}) \right ]_{ij}

\end{array}
\end{equation}

Where $D=diag(d_{1},d_{2},\cdots,d_{n})$ is the degree matrix and $d_{i}=\sum_{k=1}^{N} a_{ik}$; and $a_{ik}$ is entity in $i^{th}$ row and $k^{th}$ column of matrix $A$.
The total network communicability of individual nodes gives a measure of how well each node communicates with the other nodes of the network. To assess how communication occurs efficiency across all over the network, the sum of all the individual total communicability is considered. The total network communicability can be interpreted as a global measure of the ease of diffusion and dissemination of messages across a network. For a network with weighted adjacency matrix $A$ where $N$ is number of nodes in network, the total network communicability is given by  

\begin{equation} \label{eq.8}
\begin{array}{l}
C(A)=\sum_{i=1}^{N} \sum_{j=1}^{N} C(i,j)
\end{array}
\end{equation}

The normalized total communicability can be defined as 

\begin{equation} \label{eq.9}
\begin{array}{l}
C_{n}(A)=\frac{1}{N}\sum_{i=1}^{N} \sum_{j=1}^{N} C(i,j)
\end{array}
\end{equation}

The normalized total communicability provides a global measure of how network well-connected and can be used to compare the design and structure of different networks.

In Table 1, we compare the normalized total communicability per node, per edge and also the normalized total sub-graph centrality per node for network evolved on mobile nodes in the university campus and Montreal city datasets. We consider both cases of the weighted and unweighted social graphs among mobile nodes. 
Considering the static graph of the network formed over mobile nodes  
of dataset collected from the university campus and also the Montreal city dataset, we compare the networks capacities (communicability/centrality) when their associated graphs contain those edges with social weight higher than a specific threshold. Also, these metrics are calculated and compared for preferential attachment (Barabasi-Albert) and the small world (small graph diameter) \cite{Barabasi2003} synthetic networks with the same size \footnote{ the number of nodes is equal to the number of mobile nodes in university campus/ Montreal datasets when the social weights are higher than a specific threshold}.
 In the preferential attachment synthetic model, the edges of new node connect to the already existed nodes in the network with a probability proportional to the degree of the existed nodes. Under this scenario a scale-free node degree distribution will be appeared.
Table 1 indicates that the difference among the normalized communicability per node for the weighted and unweighted graphs is so significant. Most probably the reason is that as already was observed in Figure 1, the distribution of social weights follows the Heavy-Tail trend, means that majority of nodes have small social weight (less than one) and just a few of nodes have high social weight. It causes attenuating communicability compare to the unweighted network with the same size. As result differences between the calculated metrics for the weighted and unweighted network are so significant. The differences among the normalized total communicability per node/edge and also the  total sub-graph centrality between the weighted and unweighted networks are also so significant.
Although the normalized total communicability per node/edge and also total centrality in a social weighted network formed over university campus mobile nodes are less than Montreal network, these metrics in the university campus unweighted social network are much higher than in Montreal network. The reason for this big gap is that in university campus the average number of encounters per each mobile node is much higher than Montreal dataset.
To achieve a general perspective of communicability capacity of opportunistic network formed over the mobile nodes in these two datasets, the weighted and unweighted real world graph networks are compared with synthetic Small-World and Professional attachment networks with the same size (the same number of nodes). 
We observed a big gap difference among communicability and centrality metrics among small-world and pre-attachment synthetic networks. The normalized communicability per node and edges for small-world synthetic network are same, around 7.3891. This implies that in a synthetic small-world network, the normalized communicability per node/edge is independent of the size of the network.
The highest values of communicability and centrality measures belong to the unweighted real world dataset, Synthetic pre-attachment, small-world networks and finally the weighted real world datasets, respectively.
Considering that communicability and centrality metrics in the weighted real world network is even lower than small-world network, it implies that the normalized communicability among nodes in opportunistic network formed over mobile nodes in the university campus and Montreal city dataset are very low and sparse (although  these metrics in Montreal dataset is higher than the university campus dataset) that is due to the sparseness and coarseness of social ties among mobile nodes and also heavy-tail behavior of node degree trend.

\begin{figure}
	\centering
	\includegraphics[width=1\textwidth]{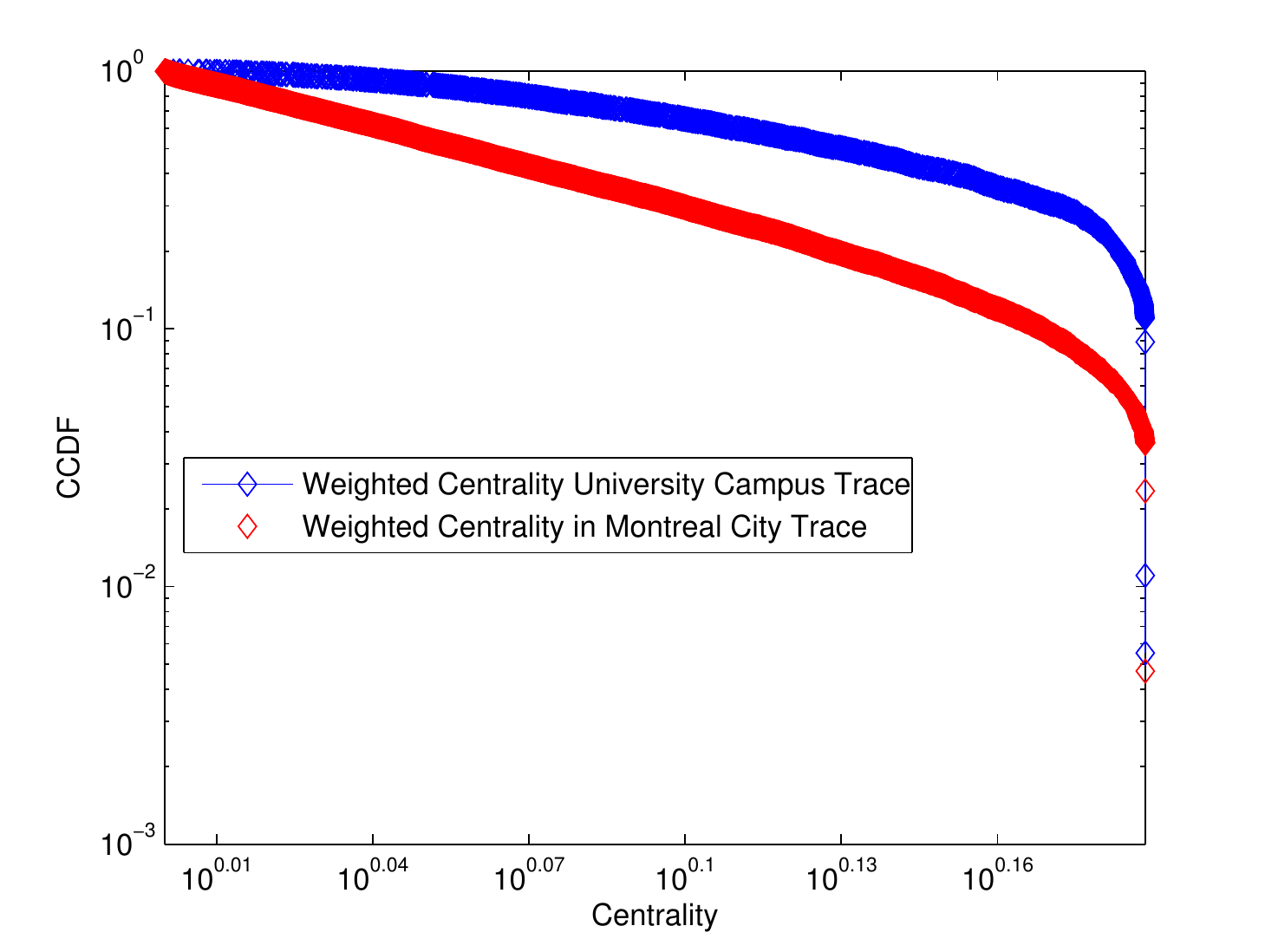}
	\caption{Compare empirical distributions of the weighted centrality in both datasets.}
	\label{fig:CompareWeightedCentrality}
\end{figure} 

\begin{figure}
	\centering
	\includegraphics[width=1\textwidth]{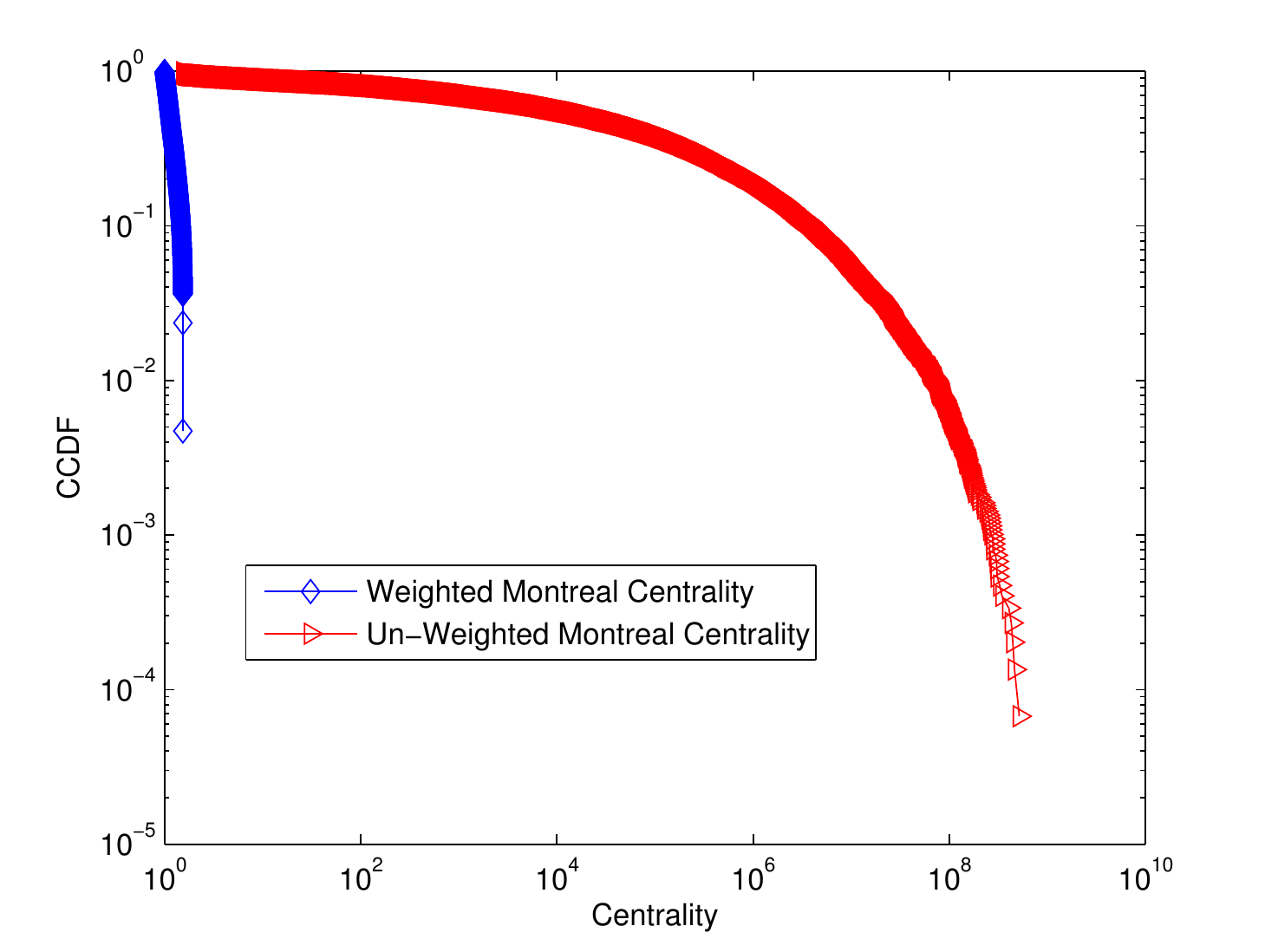}
	\caption{ The empirical distributions of the unweighted and weighted centrality in Montreal city dataset}
	\label{fig:Weighed-un-Weighted}
\end{figure} 

\begin{figure}
	\centering
		\includegraphics[width=1\textwidth]{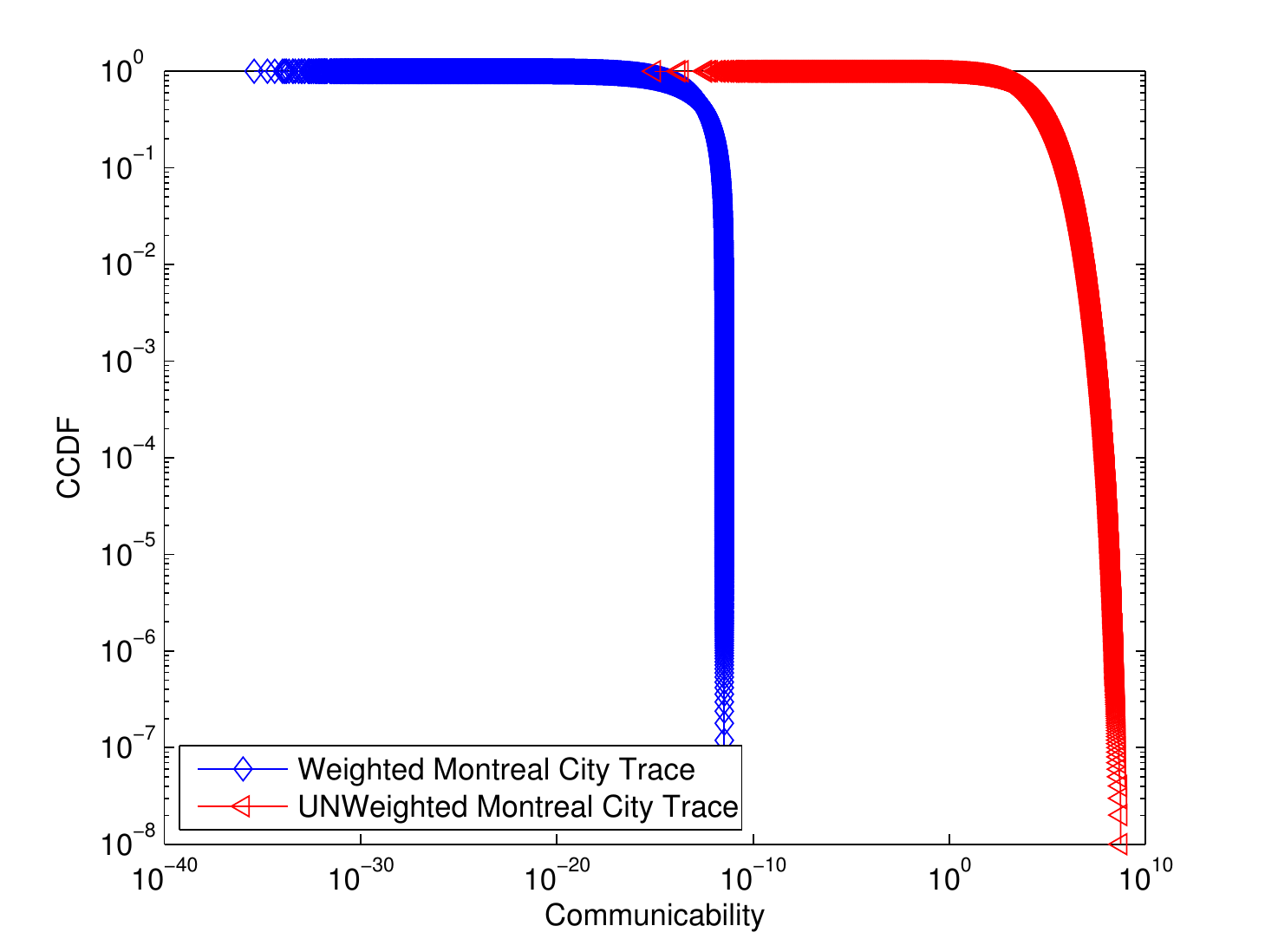}
	\caption{Compare Weighted and Unweighted communicability distributions in Montreal city dataset}
	\label{fig:Weighted-unweightedMontreal}
\end{figure}

We also show the empirical distributions of weighted and unweighted communicability/centrality of the university campus and Montreal city area in Figures \ref{fig:CompareWeightedCentrality} and \ref{fig:Weighed-un-Weighted}. As we already observed in Table 1, there is a wide gap among weighted and unweighted centrality and also communicability in both datasets. Also, the distributions imply that communicability/centrality among the majority of nodes is so sporadic rather to the minority of nodes. Also the distributions of centrality measure in Figure \ref{fig:CompareWeightedCentrality} implies that nodes in the university campus have the more important role rather to the nodes in Montreal datasets since nodes density in university campus is much higher than Montreal dataset. In Figure \ref{fig:Weighted-unweightedMontreal} the empirical distributions of weighted and unweighted communicability of Montreal dataset depicted which implies so significant big gap differences of communicability for these two cases. On the other hand in each category (weighted/unweighted), distribution of node communicability is so heterogeneous. It means the communicability among the majority of nodes is low and sporadic while among the minority of node is high.

\begin{figure}
	\centering
	\includegraphics[width=1\textwidth]{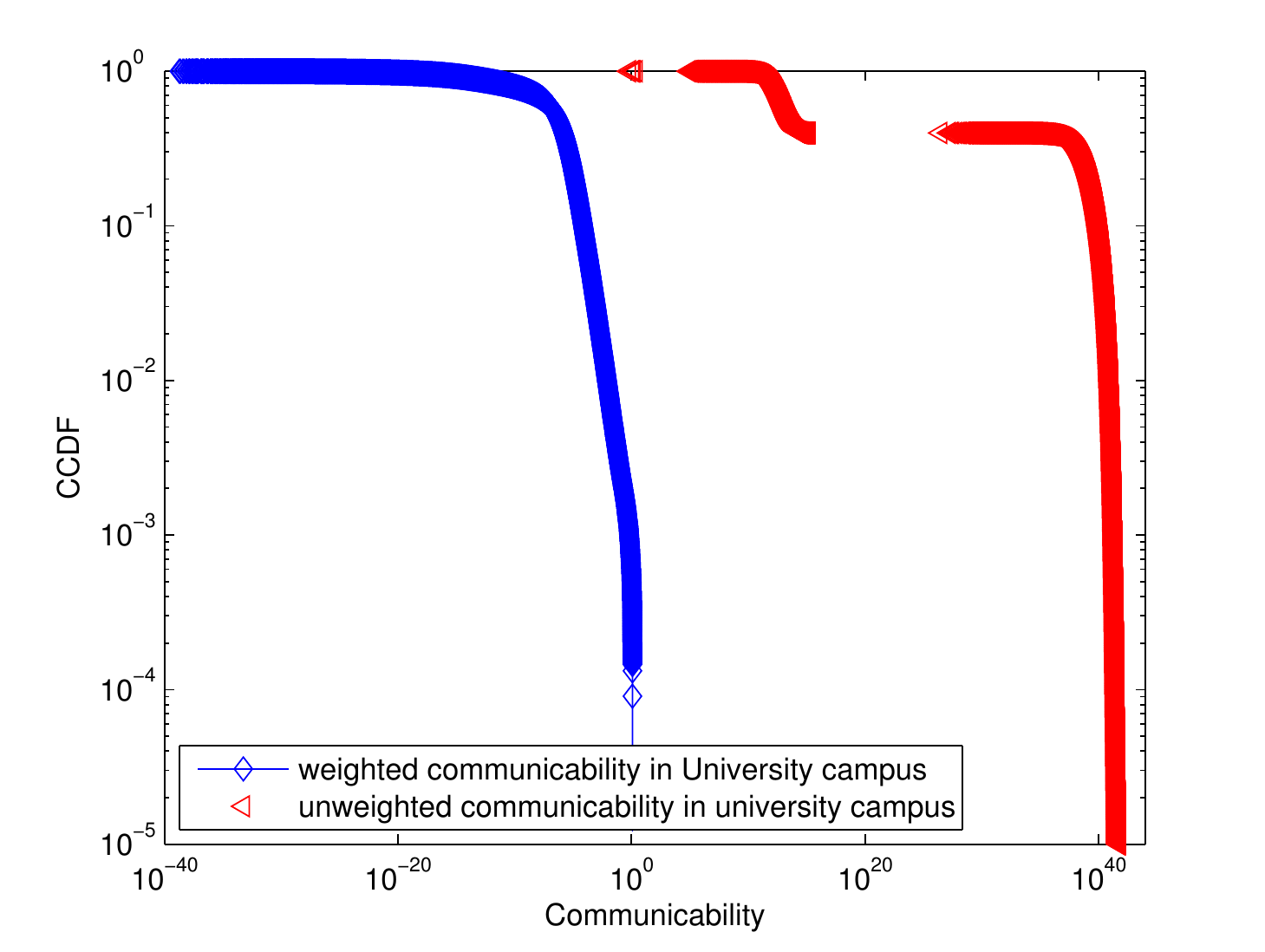}
	\caption{Compare empirical Unweighted and Weighted Communicability distributions in University Campus dataset}
	\label{fig:Weighted-unweightedunicampus}
\end{figure} 

Figure \ref{fig:Weighted-unweightedunicampus} indicates distributions of weighted and unweighted of communicability measure of university campus which again indicates significant differences between the weighted and unweighted communicability cases. Addition to the wide span range of the unweighted communicability measure, we observe even a discontinuously in the distribution of communicability measure.  
On the other hand, we observe span range of communicability measure in university campus much wider than Montreal dataset, which implies a higher average of communicability per nodes in the university campus and as result diffusion of data at the university campus environment.

\begin{figure}
	\centering
	\includegraphics[width=1\textwidth,trim={1cm 10cm 0 3cm},clip]{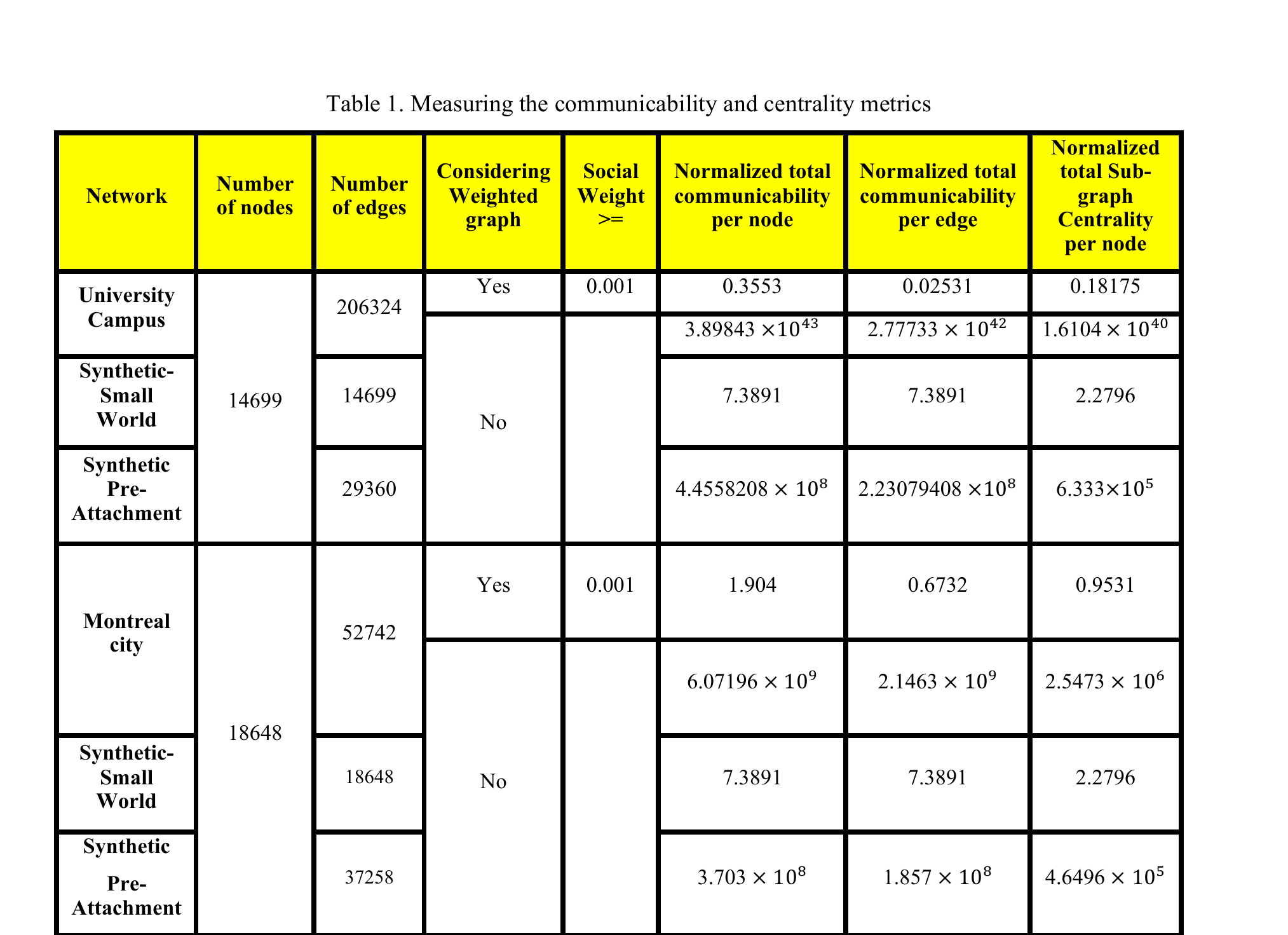}
	{\captionsetup{labelformat=empty}}
	\label{}
\end{figure} 

\section{Temporal Network}

In this paper we analysis temporal metrics that can be used to achieve a more realistic approximation of network capacity measure and also the speed (delay) of information diffusion in DTN and opportunistic networks by considering the evolution of a network from a global view.
We study how centrality and communicability metrics are able to capture the temporal characteristics of time-varying graphs compared to the metrics used in the past on static graphs. By our knowledge, the temporal dimension of these social interactions has often been neglected or underestimated while developing analytical methods for social and complex network analysis.
Since static graphs aggregate all links as appearing at the same time, they do not capture key temporal characteristics such as duration of contacts, ICT, recurrent contacts and time order of contacts along a path. For this reason, they may give us an inaccurate estimate of the potential paths connecting pairs of nodes and they can not provide reliable estimation about the communicability associated with the information spreading process. One the other hand the static metrics are not sufficient where temporal information is inherent in the network and relying on static metrics just giving a too coarse-grained view in networks while the temporal dynamics is an essential component of the phenomenon under observation such as human interactions over time.
We analysis the Temporal Communicability which can be used to assess how fast and easy information spread to all nodes by means of transitive connections between them.

\subsection{Temporal Graph}

A graph $G_{t}$ obtained by aggregating all the contacts appearing in a given interval $ \left [t,t+\Delta t \right ]$ represents the state of the system in that interval, i.e., it is a snapshot which captures the configuration of the links at that particular time Interval.
If we consider a sequence of successive non-overlapping time-window (snapshot) $ \left [ t_{m}, t_{m}+\Delta t_{m} \right]$; $i=1,\cdots, M$; then we obtain a time-varying graph, which is the simplest graph representation of a set of contacts that takes into account their duration and their temporal ordering.
So a time varying graph can be defined as an ordered sequence of $M$ graphs ${G_{1}, G_{2},\cdots, G_{M}}$ defined over $N$ nodes, where each graph $G_{i}$ in the sequence represents the state of the network, i.e., the configuration of links, in the snapshot $\left[t_{i},t_{i}+\Delta t\right]$, $i = 1,\cdots, M$.
In this notation, the $t_{M}+\Delta t_{M}-t_{1}$ is the temporal length of the observation period $T$. 
By choosing different $∆t_{i}$, it is possible to obtain different representations of systems at different temporal scale. In case that $\Delta t_{i}\rightarrow 0$, an infinite sequence of graphs will be achieved where each graph correspond to the configuration of contact at given instance $t_{i}$. 
This sequence of graphs may contain a certain number of empty graphs corresponding to the period in which no contact recorded and in the case that $\Delta t_{i}=T$, the time-varying graph will be changed to aggregated static graph where all temporal information will be lost.
Suppose $ A^{\left [ k \right ]} $ denote the adjacency matrix for the network in $k^{th}$ snapshot. We consider links as undirected and unweighted and avoiding self-loop so $A_{ii}^{\left [ k \right ]}\equiv 0$.
Given two node $i$ and $j$, the shortest temporal distance $d_{ij}$ is defined as shortest temporal path length. This can be considered as the required number of snapshots (or temporal hops) for information to be spread from node $i$ to node $j$.

\subsection{Snapshot Window Size}

One of the challenges for defining time varying graph is choosing the snapshot duration $\Delta t_{i}$ which depends on the temporal granularity of collected dataset and also affects the capturing dynamics of network. Assuming given oversampled observation of a dynamic network, the aim is that to find correct temporal resolution at which meaningful information about the structure of the network is revealed.
By choosing different snapshot window sizes, it is possible to obtain the different representation of the network at different temporal scales such as hourly, daily, weekly and monthly.
If $\Delta t_{i}$ is too large, the aggregated network over $\Delta t_{i}$ will not capture a lot of the critical temporal information such as edge concurrency and time propagating path. As result the time series graph $G_{t}$ can not correctly represent the structural variation on the network. On the other hand, if $\Delta t_{i}$ is too small, the dynamic network is aggregated over insufficient time, interesting phenomena such as the formation of component and clusters might not be evident meanwhile in this situation higher computation load will be imposed on processors. Several mechanisms for selecting the size of snapshot windows have been proposed \cite{Hossmann2009}.
An ideal solution for choosing $\Delta t_{i}$ is considering an adaptive variable $\Delta t_{i}$ which decreases in time intervals that network change dynamically fast and on the other hand choosing bigger $\Delta t_{i}$ when network dynamic changes slowly. But for this goal we need feedback on network evolution over time. Meanwhile here we ignore adaptive variable $\Delta t_{i}$ and just consider a fix $\Delta t_{i}$. In our experiment, we evaluate hourly, daily, weekly and monthly snapshots.

\subsection{Analyzing Temporal Communicability}

For defining concept of temporal communicability, first of all the dynamic walk should be clearly defined. 
A dynamic walk of length $l$ from node $v_{i}$ to node $v_{l+1}$ is made of sequence of edges connecting 
$v_{i}\rightarrow v_{i+1},v_{i+1}\rightarrow v_{i+2},\cdots,v_{l}\rightarrow v_{l+1}$ and a non-decreasing of a sequence of times $t_{r_{i}}\leq t_{r_{i+1}}\leq \cdots \leq t_{r_{i+l}}$ such that $a_{v_{m},v_{m+1}}^{\left[r_{m}\right]}$ is entity of adjacency matrix in $r_{m}^{th}$ snapshot indicates availability edge between $v_{m}$ and $v_{m+1}$.
Temporal communicability can be defined according to the extension of Katz centrality in static graph \cite{Katz1953}. 

Katz centrality measures the tendency of node $i$ to interact with node $j$ \cite{Nicosia2013}. So the multiplication of adjacency matrix in different snapshots; \\
$A^{\left [ r_{i} \right ]} A^{\left [ r_{i+1} \right ]} \cdots A^{\left [ r_{i+l} \right ]}$, includes $ij$ entity that measuring the number of dynamic walks with length $l$ starting from node $v_{i}$ and end to node $v_{j}$. So the \textbf{dynamic communicability matrix} in $k^{th}$ snapshot according to the Katz centrality can be defined as:

\begin{equation} \label{eq.11}
\begin{array}{l}
C^{k}=(I-\gamma A^{\left [ 0 \right ] })^{-1} (I-\gamma A^{\left [ 1 \right ]})^{-1} \cdots (I-\gamma A^{\left [ k \right ]})^{-1}
\end{array}
\end{equation}

Where $\gamma$ should satisfy $\gamma<\frac{1}{max_{k}(\phi( A^{\left [ k \right ]}))}$ condition, while the $\phi(.)$ is called spectral radius and is largest eign-values of adjacency matrix. The identity Matrix $I$ in equation (10), for message spreading in DTN and opportunistic network, allows that message waits in nodes until a new connection appear in next snapshots. So here $C_{ij}^{k}$ measures how well data message can be passed from node $i$ to node $j$.

Then total communicability in temporal network can be defined 

\begin{equation} \label{eq.12}
\begin{array}{l}
C_{t}=\sum_{i=1}^{N} \sum_{j=1}^{N} C_{ij}^{k}
\end{array}
\end{equation}

Since temporal networks are defined on snapshots windows, the normalized total communicability per snapshot can be defined as

\begin{equation} \label{eq.13}
\begin{array}{l}
C_{ave}=\frac{C_{t}}{M}=\frac{1}{M}\sum_{i=1}^{N} \sum_{j=1}^{N} C_{ij}^{k}
\end{array}
\end{equation}

Where $M=\left [ \frac{T}{\Delta t} \right ]+1$; $T$ and $\Delta t$ are the total duration of dataset and snapshot windows, respectively. 
Table \ref{Table 2} indicates the calculated temporal total communicability for different snapshots over all duration of university campus and Montreal city datasets:

\begin{table}[]\label{Table 2}
	\centering
	\captionsetup{labelformat=empty}
\caption{Table 2: Total communicability with different snapshots}
	\resizebox{\textwidth}{!}{%
		\begin{tabular}{|l|c|c|c|c|c|}
			\hline
			\multicolumn{2}{|l|}{} & \begin{tabular}[c]{@{}c@{}}Hourly \\ snapshot\end{tabular} & \begin{tabular}[c]{@{}c@{}}Daily\\ snapshot\end{tabular} & \begin{tabular}[c]{@{}c@{}}Weekly\\ snapshot\end{tabular} & \begin{tabular}[c]{@{}c@{}}Monthly\\ snapshot\end{tabular} \\ \hline
			\multirow{3}{*}{\begin{tabular}[c]{@{}l@{}}University\\ Campus\end{tabular}} & \begin{tabular}[c]{@{}c@{}}Total \\ communicability\end{tabular} & $8.918\times 10^{17}$ & $6.6063 \times 10^{5}$ & 27313 & 22407 \\ \cline{2-6} 
			& \begin{tabular}[c]{@{}c@{}}Normalized total \\ communicability\\  per node\end{tabular} & $6.00671 \times 10^{13}$ & 44.9438 & 1.8582 & 1.5244 \\ \cline{2-6} 
			& \begin{tabular}[c]{@{}c@{}}Number \\ of snapshots\end{tabular} & 3134 & 131 & 19 & 5 \\ \hline
			\multirow{3}{*}{\begin{tabular}[c]{@{}l@{}} Montreal \\ city \end{tabular}} & \begin{tabular}[c]{@{}c@{}}Total \\ communicability\end{tabular} & \multicolumn{1}{l|}{} & \multicolumn{1}{l|}{$2.3839\times 10^{21}$} & \multicolumn{1}{l|}{$2.8268\times 10^{5}$} & \multicolumn{1}{l|}{$2.427\times 10^{4}$} \\ \cline{2-6} 
			& \begin{tabular}[c]{@{}c@{}}Normalized \\ Communicability \\ per node\end{tabular} & \multicolumn{1}{l|}{} & \multicolumn{1}{l|}{$1.278367\times 10^{17}$} & \multicolumn{1}{l|}{15.1578} & \multicolumn{1}{l|}{ 1.30 } \\ \cline{2-6} 
			& \begin{tabular}[c]{@{}c@{}}Number of \\ Snapshot\end{tabular} & 25675 & 1070 & 153 & 36 \\ \hline
		\end{tabular}%
	}
\end{table}

\begin{figure}
 	\centering
 	\includegraphics[width=1\textwidth]{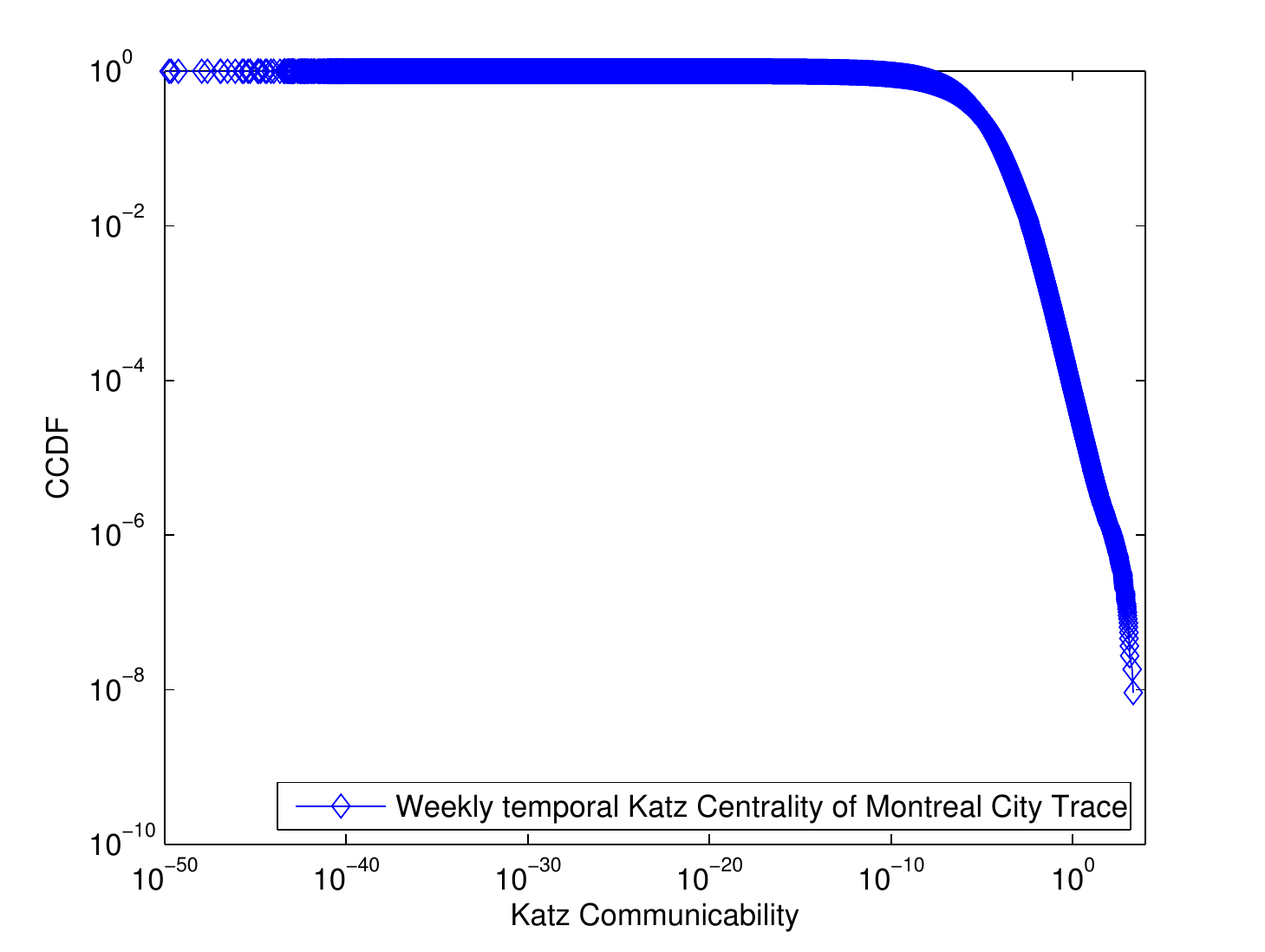}
 	\caption{Weekly Temporal Katz Communicability }
 	\label{fig:WeeklyTemporalKatzCommunicabilityMontreal}
 \end{figure}

\begin{figure}
  	\centering
  	\includegraphics[width=1\textwidth]{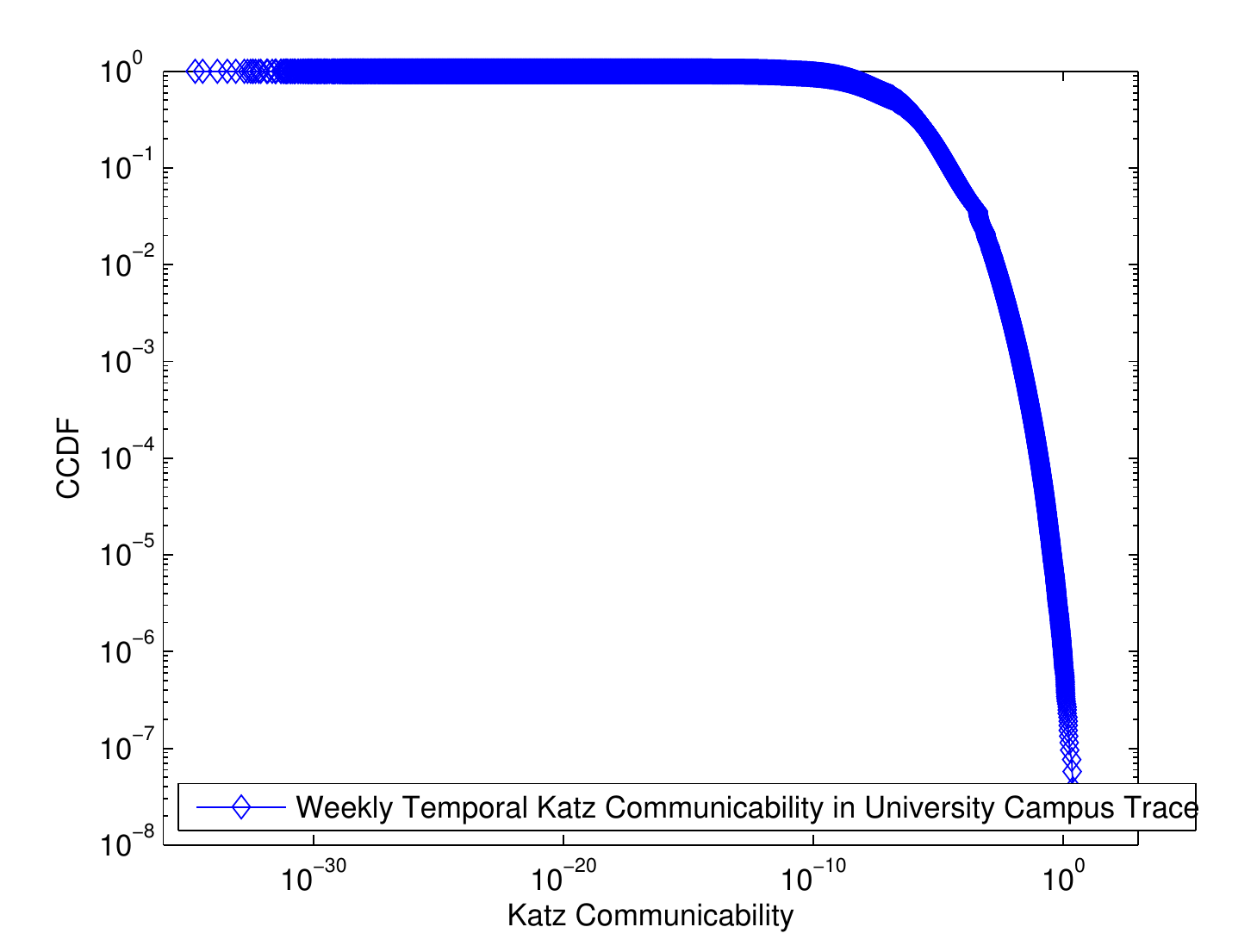}
  	\caption{Weekly Temporal Katz Communicability}
  	\label{fig:WeeklyTemporalKatzCommunicabilitycampusUniversity}
  \end{figure}

To calculate temporal communicability in equation (12), we 
 choose \\ $\gamma$=$\dfrac{0.85}{max_{k}(\phi(\left [ A^{k} \right ]))}$. The number of nodes in temporal graph of university campus and Montreal city is 14699 and 18648, same size of associated static networks.
Table 2 indicates that total communicability decreases by increasing snapshot windows. There are significant decreasing between Hourly and Daily and also Daily and Weekly snapshots. 
Significant gap between average of total communicability per node for temporal networks (for short snapshot window) and relevant unweighted static networks with same size  were observed, while this difference decreases for larger snapshot windows of temporal Network. This observation confirms that using static graph for modeling dynamic network can be very overestimating and unrealistic.
 In Figure \ref{fig:WeeklyTemporalKatzCommunicabilityMontreal} and \ref{fig:WeeklyTemporalKatzCommunicabilitycampusUniversity}, the distributions of temporal Katz communicability for weekly snapshot window for Montreal city area and university campus environments, respectively. In both distributions, communicability among majority of mobile nodes is very low and for minority of them is high. Temporal Katz communicability in Montreal city area cover wider range rather university campus.

\section{Conclusion}

In this paper, we analyzed the communicability and centrality metrics for the opportunistic network formed over mobile nodes on a university campus and some spot points in Montreal city area. We analyzed both static and dynamic graph networks.
In the static case, synthetic and real world trace networks with same sizes analyzed and we observed a significant big gap among communicability/centrality measure among synthetic and unweighted real world network with the same size. We also observed big a gap among the calculated communicability and centrality metrics among weighted and unweighted real world networks formed on mobile nodes in real world environment that are due to the heavy-tail distribution of the social weight strength tie among mobile nodes. It means that since for the majority of mobile nodes social weights are very low and less than one, it causes attenuating communicability compared to the unweighted network with the same size. The calculated metrics for weighted real world networks are lowest, implies communicability capacity is low which is due to the sparseness and coarseness of social weights among mobile nodes.
For temporal networks, we evaluated total communicability over Hourly, Daily, Weekly and Monthly snapshots. We observed that by increasing the size of snapshot windows, the total communicability decreases, especially for the hourly and weekly snapshots, decreasing is more significant. It implies that calculation of communicability metric in the temporal network is very dependent on the size of snapshot window. On the other hand, the difference between total communicability in unweighted static and temporal graph networks with the same size is significant even for short snapshot windows such as daily and weekly. It implies that modeling the dynamic network with static graph may cause to unrealistic or even mislead results.

\bibliographystyle{spmpsci}

\bibliography{matteoBib}

\end{document}